\documentstyle[12pt]{article}

\newlength{\dinwidth}
\newlength{\dinmargin}
\newcommand{\resection}[1]{\setcounter{equation}{0}\section{#1}}

\begin{document}

\font\unige=cmbx10 scaled\magstep3
\font\schola=cmr10 %scaled\magstep2
\centerline{{\unige UNIVERSIT\'E DE GEN\`EVE}}
\smallskip
\centerline{{\schola {SCHOLA GENEVENSIS MDLIX}}}
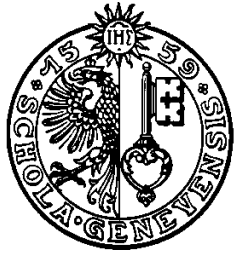
\goodbreak
\begin{center}
\bigskip\vbox
{
\vskip 3truecm
\noindent
\includegraphics{unige.ps}
\vskip 2truecm
\noindent
}
\end{center}
\bigbreak
\begin{center}
  \begin{Large}
  \begin{bf}
SYMMETRIES FOR VECTOR AND AXIAL-VECTOR MESONS$^*$\\
  \end{bf}
  \end{Large}
  \vspace{1.2cm}
  \begin{large}
R. Casalbuoni$^{(a,b)}$, A. Deandrea$^{(c)}$, S. De Curtis$^{(b)}$\\
D. Dominici$^{(a,b)}$, F. Feruglio$^{(d)}$, R. Gatto$^{(c)}$,
M. Grazzini$^{(b)}$\\
  \end{large}
\vspace{0.7cm}
$(a)$ Dipart. di Fisica, Univ. di Firenze, Largo E. Fermi, 2 - 50125
Firenze (Italy)\\
$(b)$ I.N.F.N., Sezione di Firenze, Largo E. Fermi, 2 - 50125
Firenze (Italy)\\
$(c)$ D\'ept. de Phys. Th\'eor., Univ. de Gen\`eve, CH-1211 Gen\`eve 4
(Switzerland)\\
$(d)$ Dipart. di Fisica, Univ. di Padova, Via Marzolo, 8 - 35131
Padova (Italy)
\end{center}
\vspace{1.5cm}
\begin{center}
UGVA-DPT 1995/02-881\\
hep-ph/9502247\\
February 1995
\end{center}
\vspace{3cm}
\noindent
$^*$ Partially supported by the Swiss National Foundation\\
\newpage
\thispagestyle{empty}
\begin{quotation}
\vspace*{5cm}
\begin{center}
\begin{Large}
\begin{bf}
ABSTRACT
\end{bf}
\end{Large}
\end{center}
\vspace{1cm}
\noindent
We discuss the possible symmetries of the effective lagrangian
describing interacting pseudoscalar-, vector-, and axial-vector
mesons. Specific choices of the parameters give rise to
an $[SU(2)\otimes SU(2)]^3$ symmetry. This symmetry can be obtained either
{\it \'a la} Georgi, with all states in the spectrum remaining
massless, or in a new realization,
implying decoupling of the pseudoscalar bosons from degenerate
vector and axial-vector mesons. This second possibility, when
minimally coupled with $SU(2)\otimes U(1)$ electroweak gauge vector bosons,
seems particularly appealing. All deviations in the low-energy electroweak
parameters are strongly suppressed, and the vector and axial-vector
states turn out to be quite narrow. A nearby strong electroweak sector
would then be conceivable and in principle it could be explored with relatively
low energy accelerators.
\end{quotation}
\newpage
\newcommand{\f}[2]{\frac{#1}{#2}}
\newcommand{\be}{\begin{equation}}
\newcommand{\ee}{\end{equation}}
\newcommand{\bea}{\begin{eqnarray}}
\newcommand{\eea}{\end{eqnarray}}
\newcommand{\nn}{\nonumber}
\newcommand{\dd}{\displaystyle}
\newcommand{\ct}{c_\theta}
\newcommand{\st}{s_\theta}
\newcommand{\cdt}{c_{2\theta}}
\newcommand{\sdt}{s_{2\theta}}

\setcounter{page}{1}
\def\qq{<{\overline u}u>}
\def\slash#1{\setbox0=\hbox{$#1$}#1\hskip-\wd0\dimen0=5pt\advance
       \dimen0 by-\ht0\advance\dimen0 by\dp0\lower0.5\dimen0\hbox
  to\wd0{\hss\sl/\/\hss}}
\def\vmu{{\bf V}_\mu}
\def\amu{{\bf A}_\mu}
\def\lmu{{\bf L}_\mu}
\def\rmu{{\bf R}_\mu}
\def\rmus{{\bf R}^\mu}
\def\gs{{g''}}
\def\lq{\left [}
\def\rq{\right ]}
\def\dmu{\partial_\mu}
\def\dnu{\partial_\nu}
\def\dmus{\partial^\mu}
\def\dnus{\partial^\nu}
\def\gp{g'}
\def\gpt{{\tilde g'}}
\def\ggs{\frac{g}{\gs}}
\def\eps{{\epsilon}}
\def\tr{{\rm {tr}}}
\def\V{{\bf{V}}}
\def\W{{\bf{W}}}
\def\Wt{\tilde{\bf {W}}}
\def\Y{{\bf{Y}}}
\def\Yt{\tilde{\bf {Y}}}
\def\L{{\cal L}}
\def\s{s_\theta}
\def\c{c_\theta}
\def\gt{\tilde g}
\def\et{\tilde e}
\def\At{\tilde A}
\def\Zt{\tilde Z}
\def\Wpt{\tilde W^+}
\def\Wmt{\tilde W^-}

\resection{Introduction}

In this note we will discuss some  properties of a  low-energy  effective
theory  of light pseudoscalar mesons
and  vector and axial-vector mesons.
The obvious example of such a light bosonic spectrum
is low-energy QCD.
The other situation we have in mind, and to which the majority of our
considerations will refer to, is dynamical breaking of
the electroweak
symmetry, with Goldstone bosons absorbed by the massive
gauge vector bosons $W$ and $Z$ and a set of vector and
axial-vector resonances.
The effective theory can be built on the basis of the
low-energy symmetry properties, using the well known techniques
of CCWZ \cite{ccwz}, considering the pseudoscalar mesons as Goldstone
bosons of a spontaneously broken $G$ group and the vector and axial-vector
mesons
as matter fields transforming under the unbroken subgroup $H$ of $G$.

It has also appeared convenient at least formally, to describe the vector
fields
as gauge bosons of a spontaneously broken local symmetry $H'$,
the so-called hidden symmetry \cite{bando}\cite{bala}. In this description,
the vector mesons acquire masses via a Higgs mechanism,
by eating the would-be Goldstone bosons related to the breaking
of $H'$. The explicit presence of these modes,
which are absent in the traditional CCWZ construction, is the
peculiar feature of the hidden gauge symmetry description.
Within this approach, the symmetry group of the theory is
enlarged to $G'=G\otimes H'$, with $G$ and $H'$ acting
globally and locally, respectively. The invariance group of
the vacuum is $H_D$, the diagonal subgroup of $H\otimes H'$,
($H'\supseteq H$), formally isomorphic to $H$.

Here we will focus our attention
to the specific case: $G=SU(2)_L\otimes SU(2)_R$,
$H'=SU(2)_L\otimes SU(2)_R$ and $H_D=SU(2)_V$, the latter being
the diagonal $SU(2)$ subgroup of $G'$.
The spontaneous breaking of $G'$ down to $H_D$ gives rise to
nine Goldstone bosons. Six of them are eaten up by the vector and
axial-vector mesons, triplets of $SU(2)_V$. The other three
remain in the spectrum as massless physical particles  (unless
a part of the group $G$ is promoted to a local symmetry).
This particular situation has been discussed in
ref. \cite{bando} for the QCD case and in ref. \cite{assiali} in
the context of dynamical electroweak symmetry breaking.

In the present note we would like to study in more detail the symmetry
properties of the effective theory, pointing out that,
in particular cases, the overall symmetry can be larger
than the one requested by construction, namely that associated
to the symmetry group $G'$. We shall find that a maximal symmetry
$[SU(2)\otimes SU(2)]^3$ can be realized in the low energy effective lagrangian
for pseudoscalar-, vector-, and axial-vector mesons for particular choices of
the parameters. One choice naturally generalizes the vector-symmetry of Georgi,
when axial-vector mesons are also included. This choice might offer some
starting point for QCD calculations, but its eventual usefulness remains to be
seen. A second choice will be the main object of our investigation. It may be
useful in relation to strong electroweak breaking schemes and it has the
peculiarity of allowing for a low energy strong electroweak resonant sector, at
the same time satisfying the severe present experimental constraints,
particularly from LEP data. In this sense such a choice appears of interest in
view of its possible testing within existing or future machines of relatively
low energy. These questions will be discussed below.

\resection{Vector-Axial Symmetries}

We recall that the nine Goldstone bosons
can be described by three independent
$SU(2)$ elements: $L$, $R$ and $M$, whose transformation properties
with respect to $G$ and $H'$ are the following
\be
L'= g_L L h_L,~~~~~~R'= g_R R h_R,~~~~~~M'= h_R^\dagger M h_L
\ee
where $g_{L,R}\in G$ and $h_{L,R}\in H'$. Beyond the invariance
under $G'$, we will require also an invariance under the following
discrete left-right transformation, denoted by $P$
\be
P:~~~~~~~L\leftrightarrow R,~~~~~M\leftrightarrow M^\dagger
\ee
which also ensures that the low-energy theory is parity conserving.

Ignoring for a moment the transformation laws of eq. (2.1),
the largest possible global symmetry of the low-energy theory
is determined by the request of maintaining for the transformed
variables $L'$, $R'$ and $M'$, the character of $SU(2)$ elements.
This selects as maximal symmetry the group $G_{max}=[SU(2)\otimes
SU(2)]^3$, consisting of three independent $SU(2)\otimes SU(2)$
factors, acting separately on each of the three variables.
Thus, it may happen that, for specific choices of the parameters
characterizing the theory, the symmetry $G'$ gets enlarged to
$G_{max}$. In this note we will discuss two special cases
which give rise to such a symmetry enhancement.

The most general $G'\otimes P$ invariant lagrangian is given by \cite{assiali}
\be
{\cal L}={\cal L}_G+{\cal L}_{kin}
\ee
where
\be
{\cal L}_G=-\frac{v^2}{4} [a_1 I_1 + a_2 I_2 + a_3 I_3 + a_4 I_4]
\ee
\bea
I_1&=&tr[(V_0-V_1-V_2)^2]\nn\\
I_2&=&tr[(V_0+V_2)^2]\nn\\
I_3&=&tr[(V_0-V_2)^2]\nn\\
I_4&=&tr[V_1^2]
\eea
and
\bea
V_0^\mu&=&L^\dagger D^\mu L\nn\\
V_1^\mu&=&M^\dagger D^\mu M\nn\\
V_2^\mu&=&M^\dagger(R^\dagger D^\mu R)M
\eea
The covariant derivatives are defined by
\bea
D_\mu L&=&\partial_\mu L -L \lmu\nn\\
D_\mu R&=&\partial_\mu R -R \rmu\nn\\
D_\mu M&=&\partial_\mu M -M \lmu+\rmu M
\eea
and
\be
{\cal L}_{kin}=\frac{1}{\gs^2} tr[F_{\mu\nu}({\bf L})]^2+
         \frac{1}{\gs^2}  tr[F_{\mu\nu}({\bf R})]^2
\ee
where $\gs$ is the gauge coupling constant for the gauge fields $\lmu$ and
$\rmu$, and
\be
F_{\mu\nu}({\bf L})=\partial_\mu{\bf L}_\nu-\partial_\nu{\bf L}_\mu+
          [\lmu,{\bf L}_\nu]
\ee
In eq. (2.4) $v$ represents a physical scale which
depends on the particular
context (QCD, electroweak symmetry breaking...) under
investigation.

The quantities $V_i^\mu~~(i=0,1,2)$ are, by construction,
invariant under the global symmetry $G$ and covariant under
the gauge group $H'$
\be
(V_i^\mu)'=h_L^\dagger V_i^\mu h_L
\ee
Out of the $V_i^\mu$ one can build six independent quadratic invariants,
which reduce to the four $I_i$ listed above, when parity is enforced.

For generic values of the parameters $a_1,~a_2,~a_3,~a_4$, the
lagrangian ${\cal L}$
is invariant under $G'\otimes P=G\otimes H'\otimes P$.
There are however special choices
which enhance the symmetry group.
One of these is a special limit of the case $a_1=0$, $a_2=a_3$.
One has
\be
{\cal L}_G=-\frac{v^2}{4}[2~ a_2~ tr(V_0)^2+ a_4~ tr(V_1)^2+ 2~ a_2~ tr(V_2)^2]
\ee
Although the above choice of parameters has diagonalized the
lagrangian in the variables $L$, $R$, $M$ (see eq. (2.6)), this is not yet
sufficient to modify the overall symmetry, due to the local
character of the group $H'$ which leads to a non-trivial
dependence on the gauge fields $\lmu$ and $\rmu$.
However, if we consider the limit when these gauge interactions
are turned off,
%by sending the gauge coupling $\gs$ to zero,
then the lagrangian becomes
\be
{\cal L}_G=\frac{v^2}{4}\{2~a_2~ [tr(\partial_\mu L^\dagger \partial^\mu L)+
                          tr(\partial_\mu R^\dagger \partial^\mu R)]+
                         a_4~ tr(\partial_\mu M^\dagger \partial^\mu M)\}
\ee
which is manifestly invariant under $G_{max}=[SU(2)\otimes SU(2)]^3$.
This limit describes a set of nine massless, degenerate Goldstone bosons.
It could be meaningful for vector and axial-vector mesons much
lighter than the chiral breaking scale $4 \pi v$ and almost degenerate
with the pseudoscalar particles. This case can be considered as a
generalization
of the vector symmetry described by Georgi \cite{georgi}.

Another case of interest is provided by the choice:
$a_4=0$, $a_2=a_3$.
To discuss the symmetry properties it is useful to observe
that the invariant $I_1$ could be re-written as follows
\be
I_1=-tr(\partial_\mu U^\dagger \partial^\mu U)
\ee
with
\be
U=L M^\dagger R^\dagger
\ee
and the lagrangian becomes
\be
{\cal L}_G=\frac{v^2}{4}\{a_1~ tr(\partial_\mu U^\dagger \partial^\mu U) +
                         2~a_2~ [tr(D_\mu L^\dagger D^\mu L)+
                          tr(D_\mu R^\dagger D^\mu R)]\}
\ee
Each of the three terms in the above expressions
is invariant under an independent $SU(2)\otimes SU(2)$
group
\be
U'=\omega_L U \omega_R^\dagger,~~~~~~L'= g_L L h_L,~~~~~~R'= g_R R h_R
\ee
Moreover, whereas these transformations act globally on the $U$
fields, for the variables $L$ and $R$, an $SU(2)$ subgroup is
local. Different from the previous case, there is no need of
turning the gauge interactions off. The overall symmetry is still
$G_{max}=[SU(2)\otimes SU(2)]^3$, with a part $H'$ realized as a gauge
symmetry.

The field redefinition from the variables $L$, $R$ and $M$ to
$L$, $R$ and $U$ has no effect on the physical content of the theory.
It is just a reparametrization of the scalar manifold.

The extra symmetry related to the independent transformation
over the $U$ field, could be also expressed in terms of the original
variable $M$. Indeed the lagrangian of eq. (2.4), for $a_4=0,~a_2=a_3$,
possesses the additional invariance
\be
L'=L,~~~~~~R'=R,~~~~~~M'=\Omega_R M \Omega_L^\dagger
\ee
with
\be
\Omega_L=L^\dagger \omega_L L,~~~~~~\Omega_R=R^\dagger \omega_R R
\ee

By expanding the lagrangian in eq. (2.4) in powers of the Goldstone bosons
one finds, as the lowest order contribution, the mass terms for the vector
mesons:
\be
{\cal L}_G = -\frac{v^2}{4}[a_2~ tr(\lmu+\rmu)^2 + (a_3+a_4)~
         tr(\lmu-\rmu)^2]+\cdots
%           &=& -\frac{v^2}{4}[(a_2+a_3+a_4)~ (tr(\lmu)^2 + tr(\rmu)^2)\nn\\
%           & & +2~(a_2-a_3-a_4)~ tr(\lmu\rmus)]+\cdots~~~,
\ee
where the dots stand for terms at least linear in the Goldstone modes.
By introducing the linear combinations
\be
\vmu=\frac{1}{2} (\rmu+\lmu),~~~~~\amu=\frac{1}{2} (\rmu-\lmu)
\ee
we can rewrite
\be
{\cal L}_G=-v^2[a_2~ tr(\vmu)^2 + (a_3+a_4)~ tr(\amu)^2]+\cdots
\ee

In the following we will focus on the
case $a_4=0$, $a_2=a_3$. Then, as we see from eq. (2.19),
the mixing between $\lmu$ and $\rmu$
is vanishing, and the two states are degenerate in mass.
Moreover, as it follows from eq. (2.15), the longitudinal modes
of the $\lmu$ and $\rmu$ (or $\vmu$, $\amu$) fields are entirely
provided by the would-be Goldstone bosons in $L$ and $R$. This means
that the pseudoscalar particles remaining as physical states in the
low-energy spectrum are those associated to $U$. They may in turn
provide the longitudinal components to the $W$ and $Z$ particles,
in an effective description of the electroweak breaking sector,
or simply remain true Goldstone bosons if QCD is considered.

The peculiar feature of the limit under consideration
is that these modes are decoupled from
the vector and axial-vector mesons, as one can immediately deduce by inspecting
eq. (2.15). In QCD this would mean that the coupling $g_{\rho \pi\pi}$
vanishes in the limit considered. The same remark also applies to
the $g_{\rho a_1\pi}$ coupling, which is absent.
All this appears to be rather far from the real QCD picture.
Nonetheless, it may not be unconceivable to model the low-energy
strong interactions as a perturbation around the very crude
approximation provided by the above scenario. The starting point
would correspond to a symmetry, whose breaking could be systematically
investigated by adding symmetry violating terms of increasing importance,
in a chiral expansion. It would be of course of great interest to
relate this symmetry to some particular limit of QCD.

\resection{Consequences for electroweak breaking}

Let us now consider the coupling of the model to the electroweak
$SU(2)_L\otimes U(1)_Y$ gauge fields as obtained via the minimal
substitution
\bea
D_\mu L &\to& D_\mu L+ {\bf W}_\mu L\nn\\
D_\mu R &\to& D_\mu R+ {\bf Y}_\mu R\nn\\
D_\mu M &\to& D_\mu M
\eea
where
\bea
{\bf W}_\mu &=& \f{i}{2} g \tau_a W_\mu^a\nn\\
{\bf Y}_\mu &=& \f{i}{2} \gp \tau_3 Y_\mu\nn\\
{\bf L}_\mu &=& \f{i}{2} \f{\gs}{\sqrt{2}} \tau_a L_\mu^a\nn\\
{\bf R}_\mu &=& \f{i}{2} \f{\gs}{\sqrt{2}} \tau_a R_\mu^a
\eea
with $a=1,2,3$, and by introducing
the canonical kinetic terms for $W_\mu^a$ and $Y_\mu$.
The factor $\gs/\sqrt{2}$ in the definition of
$\lmu$ and $\rmu$ comes from the re-scaling of the gauge fields
to get canonical kinetic terms (see eq. (2.8)).
The mass term for the vector mesons reads:
\be
{\cal L}_G^{(2)} = -\frac{v^2}{4}[a_1~ tr({\bf W}_\mu-{\bf Y}_\mu)^2
               + 2a_2~(tr(\lmu-{\bf W}_\mu)^2
            +tr(\rmu-{\bf Y}_\mu)^2)]
\ee
The first term in the previous equation reproduces precisely the mass term
for the ordinary gauge vector bosons in the SM, provided we
identify the combination $v^2 a_1$ with $1/(\sqrt{2} G_F)$, $G_F$
being the Fermi constant. Indeed, it is natural to think about
the model we are considering as a perturbation around the SM picture.
The SM relations are obtained in the limit $\gs \gg g,g'$. Actually,
for a very large $\gs$, the kinetic terms for the fields $\lmu$ and $\rmu$
drop out, and ${\cal L}_G^{(2)}$ reduces to the first term in eq. (3.3).
In the following we will assume:
\be
a_1=1~~~,~~~~~ v^2=1/(\sqrt{2} G_F)
\ee
By writing  ${\cal L}_G^{(2)}$ in terms of the charged and the neutral
fields  one finds:
\bea
{\cal L}_G^{(2)} &=& \frac{v^2}{4}[(1+2 a_2)g^2 W_\mu^+ W^{\mu -}+
                      a_2 \gs^2 (L_\mu^+ L^{\mu -}+R_\mu^+ R^{\mu -})\nn\\
                & &-\sqrt{2}a_2 g\gs (W_\mu^+ L^{\mu -}+W_\mu^- L^{\mu +})]
\nn\\
&+& \frac{v^2}{8}[(1+2 a_2) (g^2 W_3^2+\gp^2 Y^2)+
                      a_2 \gs^2 (L_3^2+R_3^2)\nn\\
                & &- 2 g \gp W_{3\mu}Y^\mu
                -2 \sqrt{2}a_2\gs (g W_3 L_3^\mu +\gp Y_\mu R_3^\mu)]
\eea

In the charged sector the fields $R^\pm$ are completely decoupled
 from the remaining states, for any value of $\gs$.
Their mass is given by:
\be
M^2_{R^\pm}=\f{v^2}{4} a_2 \gs^2
\ee
On the contrary the modes $L^\pm$ and $W^\pm$ are mixed. By denoting
with ${\hat L}^\pm$ and ${\hat W}^\pm$ the mass eigenstates, one
obtains the following masses:
\be
M^2_{{\hat W}^\pm}=\f{v^2}{4} g^2[1-2(\ggs)^2+\cdots]~~~,~~~~~
M^2_{{\hat L}^\pm}=\f{v^2}{4} a_2 \gs^2 [1+2 (\ggs)^2+\cdots]
\ee
where the dots stand for higher order terms in $g/\gs$.
The physical particle ${\hat L}^\pm$ is a combination of $L^\pm$
and $W^\pm$, which, for
small values of $g/\gs$, is mainly oriented along the $L^\pm$ direction.

In the neutral sector there is, as expected, a strictly massless combination
which corresponds to the physical photon associated to the unbroken
$U(1)$ gauge group. The remaining states, here denoted by
${Z}$, ${L}_0$ and ${R}_0$, are massive and
essentially aligned along the combinations $(\ct~ W_3-\st~ Y)$,
$L_3$ and $R_3$ respectively
($\st$ and $\ct$ denote the sine and the cosine
of the Weinberg angle $\theta$, with $\tan\theta=g'/g$).
Unlike the charged case, however,
the physical state $R_0$ is not completely decoupled, in fact at the
leading order in $g/\gs$,
it possesses a tiny component along the
$Y$ direction. The $L_0$ state has in turn a small contribution from the
$W_3$ field.
The non-vanishing masses are given by:
\bea
M^2_Z &=& \f{v^2}{4} \f{g^2}{\ct^2}[1-\f{1+\cdt^2}{\ct^2}(\ggs)^2
+\cdots]\nn\\
M^2_{{L_0}} &= & \f{v^2}{4} a_2 \gs^2 [1+2(\ggs)^2+\cdots]\nn\\
M^2_{{R_0}} &= & \f{v^2}{4} a_2 \gs^2 [1+2\f{\st^2}
    {\ct^2}(\ggs)^2+\cdots]
\eea
where the dots represent higher order terms in the $g/\gs$ expansion.

A non-trivial consequence of the mass spectrum of the model
and of the structure of the bilinear terms given in eq. (3.5)
concerns the low-energy effects which can be tested using the results
of the available high-precision measurements. It has become customary
to cast such analysis in the framework of the so-called
$\epsilon$ variables \cite{alt} \cite{fri}.
Alternatively one could use the analysis of Peskin and Takeuchi \cite{S}.
We recall that the
deviations from the SM expectations for $\epsilon$ parameters
are given by \cite{fri}
\bea
\delta\epsilon_1&=&e_1-e_5\nn\\
\delta\epsilon_2&=&e_2-\st^2 e_4 -\ct^2 e_5\nn\\
\delta\epsilon_3&=&e_3+\ct^2 e_4-\ct^2 e_5
\label{a16}
\eea
where we have kept into account the fact that in our case
there are no vertex or box corrections to four-fermion processes, with
the definitions
\bea
e_1&=&\frac{A_{33}-A_{WW}}{M_W^2}\nn\\
e_2&=&F_{WW}(M_W^2)-F_{33}(M_Z^2)\nn\\
e_3&=&\frac{\ct}{\st} F_{30}(M_Z^2)\nn\\
e_4&=&F_{\gamma\gamma}(0)-F_{\gamma\gamma}(M_Z^2)\nn\\
e_5&=&M_Z^2 F'_{ZZ}(M_Z^2)
\label{a17}
\eea
The quantities $A_{ij}$ and $F_{ij}$ are related to the
two-point vector boson functions
%$-i\Pi^{\mu\nu}_{ij}(q)~~~(i,j=0,1,2,3)$ are decomposed as follows:
\be
\Pi^{\mu\nu}_{ij} (q) = \Pi_{ij}(q^2) g^{\mu\nu} + (q^\mu q^\nu ~~{\rm terms})
{}~~~~~(i,j=0,1,2,3)
\label{a0}
\ee
by the decomposition:
\be
\Pi_{ij}(q^2)=A_{ij} + q^2 F_{ij}(q^2)
\label{a1}
\ee
In our case we get \cite{self}
\bea
e_1&=&0\nn\\
e_2&=&-\st^2\left(1-\f{M_Z^2}{M^2}\right)
\left(1-\f{M_Z^2\ct^2}{M^2}\right)^{-1}X\nn\\
e_3&=&0\nn\\
e_4&=&-\f{\sdt^2}{2\ct^2}\left(1-\f{M_Z^2}{M^2}\right)X\nn\\
e_5&=&\f{\ct^4+\st^4}{\ct^2}X
\eea
where
\be
X= 2 \f{M_Z^2}{M^2} (\dd\ggs)^2 \f{1}{(1-\dd\f{M_Z^2}{M^2})^2}
\ee
In the previous expressions $M^2$ denotes $v^2 a_2 \gs^2/4$, the common
squared mass of the vector bosons $\lmu$ and $\rmu$,
at leading order in the $g/\gs$ expansion.
In an expansion in $M_Z^2/M^2$, the leading terms for the corrections to the
epsilon parameters are then given by
\bea
\delta\epsilon_1&=&-\f{\ct^4+\st^4}{\ct^2}~ X\nn\\
\delta\epsilon_2&=&-\ct^2~ X\nn\\
\delta\epsilon_3&=&-X
\eea
with $X\approx 2(M_Z^2/M^2)(g/\gs)^2$.
Notice that the deviations $\delta\epsilon$'s are negative.
They are all of order $X$, which contains a double suppression
factor: $M^2_Z/M^2$ and $(g/\gs)^2$.
The sum of the SM contributions, functions of the top and Higgs masses,
and the previous deviations has to be compared with the experimental
values for the $\epsilon$ parameters, determined from all the available
low-energy data
\cite{altarelli}
\bea
\epsilon_1&=&(3.5\pm 1.7)\cdot 10^{-3}\nn\\
\epsilon_2&=&(-5.4\pm 4.7)\cdot 10^{-3}\nn\\
\epsilon_3&=&(3.9\pm 1.7)\cdot 10^{-3}
\label{eps}
\eea
Notice the relatively large error in the determination of $\epsilon_2$,
mainly dominated by the uncertainty on the $W$ mass. Indeed
$\epsilon_1$ provides the most stringent
bound on the parameter $X$. Taking into account the SM value
$(\epsilon_1)_{SM}=3.8\cdot 10^{-3}$, for $m_{top}=174~GeV$ and
$m_H=1000~GeV$, we find the $90\%$ CL limit:
\be
M(GeV)\ge  1877(\ggs)
\ee
leaving room for relatively light resonances beyond the usual SM spectrum.

We recall that the very small experimental value of the $\epsilon_3$
parameter, which measures the amount of isospin-conserving virtual
contributions to the vector boson self-energies, strongly
disadvantages the ordinary technicolor schemes, for which the contribution to
$\epsilon_3$ is large and positive.
This problem could be attributed to the vector dominance
in the dispersion relation satisfied by the $e_3$ parameter (see eq. (3.10))
\cite{S}:
\be
e_3=-\frac{g^2}{4 \pi} \int_0^\infty \frac{ds}{s^2}
[Im \Pi_{VV}(s)- Im \Pi_{AA}(s)]
\ee
where $\Pi_{VV(AA)}$ is the correlator between two vector (axial-vector)
currents.
If the vector and axial-vector spectral functions are saturated
by lowest lying vector and axial-vector resonances, one has
\be
Im \Pi_{VV(AA)}(s)=-\pi g_{V(A)}^2 \delta(s-M^2_{V(A)})
\ee
where $g_{V(A)}$ parametrizes the matrix element of the vector
(axial-vector) current between the vacuum and the state $V_\mu$ ($A_\mu$),
and $M_{V(A)}$ is the vector (axial-vector) mass.
 From the previous equations, one obtains
\be
e_3=\frac{g^2}{4}\left[\frac{g_V^2}{M_V^4}-\frac{g_A^2}{M_A^4}\right]
\ee
If the underlying theory mimics the QCD behaviour, naively scaled
 from $f_\pi\simeq 93~MeV$ to $v\simeq 246~ GeV$, it follows that the
contribution to $\epsilon_3$ is unacceptably large \cite{altarelli}.
On the contrary, in the present model the approximate
degeneracy among the masses of the vector and axial-vector
states and their couplings makes $e_3$ vanish. The $\epsilon_3$
parameter is saturated by the remaining, small,
$e_4$ and $e_5$ contributions (see eq. (3.9)).

The important feature of the model under examination is that
the previous results are protected by the symmetry discussed in
section 2. On the other hand we have tacitly assumed that, in the
context of the electroweak symmetry breaking, this symmetry
is explicitly broken only by terms of electroweak strength, originating
by the minimal coupling given in eq. (3.1). Whether such an assumption
is correct or not is a dynamical issue which could be answered
only by analyzing the underlying fundamental theory. It would be of great
interest to build up explicitly a more fundamental
theory displaying in some limit
an $[SU(2)\otimes SU(2)]^3$ symmetry, broken at low-energy by the
electroweak interactions.
On a more phenomenological level we may ask whether the low-energy
phenomenology outlined above is modified by allowing breaking terms
of more general form.

Coming back to the lagrangian of eqs. (2.3-2.5), with unspecified
values of the $a_i$ parameters, one gets \cite{bessu8}:
\be
g_V=-\frac{1}{2} a_2 \gs v^2,~~~~
g_A=\frac{1}{2} a_3 \gs v^2
\ee
Therefore, at the leading order in the weak interaction
($M_V^2=a_2\gs^2v^2/4$, $M_A^2=\break (a_3+a_4)\gs^2 v^2/4$)
\be
e_3=\left(\frac{g}{\gs}\right)^2\left[1-\frac{a_3^2}{(a_3+a_4)^2}\right]
\ee
We see that $e_3=0$ is certainly guaranteed by assuming $g_V^2=g_A^2$
and $M_V^2=M_A^2$, which is the case for $a_2=a_3$, $ a_4=0$.
If we apply a perturbation to this case respecting the $[SU(2)\otimes SU(2)]^2$
symmetry of ${\cal L}_G$, we see from eq. (3.22) that it produces an effect
on $e_3$ only if $a_4\not =0$. In terms of the $\epsilon_3$ parameter the
unsuppressed part of $\delta \epsilon_3$ is controlled by $a_4$:
\be
\delta \epsilon_3\approx 2\frac{a_4}{a_3}\left(\frac{g}{\gs}\right)^2~~~.
\ee
If $a_4>0$, the deviation for the $\epsilon_3$ parameter is positive and
potentially large, since it is no more suppressed by
the double factor contained in $X$ (see eq. (3.13-3.14)),
which still protects the $\epsilon_1$ and $\epsilon_2$ parameters. This is
what happens in the QCD-scaled technicolor models \cite{bessu8}.

We conclude this section with some remarks about the decay
of the vector mesons $\lmu$ and $\rmu$.
In the present, effective, description of the electroweak symmetry-breaking,
the Goldstone bosons in $U$ become unphysical scalars eaten
up by the ordinary gauge vector bosons $W$ and $Z$.
The absence of coupling among $U$ and the states $L$ and $R$
results in a suppression of the decay rate of these
states into $W$ and $Z$. Consider, for instance, the decay of
the new neutral gauge bosons into a $W$ pair. In a minimal model, with
only vector resonances, this decay channel is largely the dominant one.
The corresponding width is indeed given by \cite{pierre43}
\be
\Gamma (V_0\to WW)=\f{G_F^2}{24\pi} \frac{M^5}{\gs^2}
\ee
and it is enhanced with respect to the partial width into a fermion
pair, by a factor $(M/M_W)^4$ \cite{pierre43}
\be
\Gamma(V_0\to {\bar f} f)\approx G_F M_W^2 (\frac{g}{\gs})^2 M~~~.
\ee
This fact is closely related to the existence of a coupling of order $\gs$
among $V_0$ and the unphysical scalars absorbed by the $W$ boson.
Indeed the fictitious width of $V_0$ into these scalars provides,
via the equivalence theorem \cite{equ}, a good approximation to
the width of $V_0$ into a pair of longitudinal $W$ and it is
precisely given by eq. (3.24).

On the contrary, if there is no direct coupling among the new gauge bosons and
the would-be Goldstone bosons which provide the longitudinal
degree of freedom to the $W$, then their partial width
into longitudinal $W$'s will be suppressed compared to the
leading behaviour in eq. (3.24), and the width
into a $W$ pair could be similar to the fermionic width.
As one can explicitly check, this entails that the trilinear couplings
between the new gauge bosons and a $W$ pair
is no longer of order $g (g/\gs)$, but at most of order $g (g/\gs)^3$.
 From the trilinear kinetic terms and the mixing among the gauge
bosons, we find, at the first non trivial
order in $(g/\gs)$:
\bea
g_{L_0WW} &=& \sqrt{2}~ \f{g}{a_2}~(\ggs)^3\nn\\
g_{L^\pm W^\pm Z} &=& \f{\sqrt{2}}{\ct} \f{g}{a_2}~(\ggs)^3\nn\\
g_{R_0WW} &=& \sqrt{2}~ \f{g}{a_2}~\tan^2\theta ~ (\ggs)^3
\eea
The $R^\pm$ have no mixing whatsoever and therefore they will be absolutely
stable as ensured by the phase invariance $R^\pm\to R^\pm\exp{(\pm i \alpha)}
$.
This means that $g_{R^\pm W^\pm Z} = 0$ at all order in $(g/\gs)$.
It may be useful to compare the couplings of $\lmu$ and $\rmu$ into
vector boson pairs with those into fermions.
If we do not introduce any direct coupling between the new vector bosons
and the ordinary fermions, the only available interaction is the one obtained
through the mixing of $W^\pm$, $W_3$ and $Y$ with the new states.
These couplings are described by the interaction lagrangian:
\begin{equation}
{\cal L}_F=-\sqrt{2} g~\dd\ggs~ J^\mu_{3L}~ {L_0}_\mu
+\sqrt{2} g~\dd\ggs~ \f{\st^2}{\ct^2} (J^\mu_{3L}~-J^\mu_{em})~{R_0}_\mu
+ g~\dd\ggs~\left[J^+_\mu {L^-}^\mu+~h.c.\right]
\end{equation}
where $J^+_\mu$ is the usual charged current, $J^\mu_{3L}$ and $J^\mu_{em}$
are the neutral currents related to the third component of the weak isospin
and to the electromagnetic charge, respectively.
Notice that also in this case the charged state $R^\pm$ remains decoupled.

By using eqs. (3.26-3.27) we get the following expressions for the widths
of the ${\bf L}_\mu$ and ${\bf R}_\mu$ bosons:
\be
\Gamma(L_0\to WW)=\Gamma(L^\pm\to W^\pm {Z^0})=
\f{\sqrt{2}G_F M^2_W}{24\pi}M\left(\f{g}{\gs}\right)^2
\ee
\be
\Gamma(R_0\to WW)=\f{\sqrt{2}G_F M_W^2}{24\pi}\tan^4\theta~
 M\left(\f{g}{\gs}\right)^2
\ee
\be
\Gamma(L_0\to \nu{\bar \nu})=\f{1}{2}\Gamma(L^+\to e^+ \nu )=
\f{\sqrt{2}G_F M^2_W}{12\pi}M\left(\f{g}{\gs}\right)^2
\ee
\be
\Gamma(R_0\to \nu{\bar \nu})=\f{\sqrt{2}G_F M_W^2}{12\pi}\tan^4\theta~
M\left(\f{g}{\gs}\right)^2
\ee
which confirm the expected behaviour, linear in the heavy mass $M$.

\resection{Conclusions}

We have discussed the symmetry properties of an effective lagrangian
describing light pseudoscalars interacting with vector and axial-vector
mesons. The maximal symmetry of such system, $[SU(2)\otimes SU(2)]^3$,
is exhibited by specific choices of the parameters characterizing
the model.

One of these choices represents the natural generalization
for including axial-vectors in
the so-called vector symmetry, proposed by Georgi in the context of the
strong interactions. In its realization the spectrum consists of a set of
nine massless Goldstone bosons, six of which representing the longitudinal
degrees of freedom of the vector and axial-vector mesons. This description
could be meaningful for light vector and axial-vector states, almost
degenerate with the pseudoscalar particles. Its eventual usefulness for
QCD remains to be investigated.

A second possible realization of the maximal symmetry, which we have
discussed in more detail, is instead
characterized by a degeneracy among the vector and axial-vector
particles which, however, are no longer bound to have the same mass of
the pseudoscalar modes. The latter, in the symmetry limit, are completely
decoupled from the vector meson states.
By minimally coupling this model to the electroweak vector bosons
$W^a_\mu$ and $Y_\mu$, we obtain an effective model for the
electroweak symmetry breaking, in which the $[SU(2)\otimes SU(2)]^3$
symmetry is broken by the weak interaction terms.
The physical features of such model are quite peculiar.
They are tightly related to the special breaking of
$[SU(2)\otimes SU(2)]^3$ considered, limited to terms
originating from the minimal coupling to the electroweak vector
bosons. Additional breaking terms, as allowed in general by the
hidden symmetry construction, would lead to substantial modifications.

In our vector-axial degenerate model all the low-energy
effects are strongly suppressed, unlike the usual technicolor-inspired
effective models. In particular, the tiny deviations
in the $\epsilon$ parameters are negative and of order $M_Z^4/M^4$,
$M$ denoting the common mass of the vector mesons. This leaves room
for relatively light vector resonances.
Moreover, due to the vanishing of the coupling between the vector mesons
and the would-be Goldstone modes absorbed by $W$ and $Z$, the decay
widths of the vector and axial-vector states are equally shared among the
vector boson pair and fermion channels.
Such widths, being of order $M_Z/M$ times the typical $Z$ fermionic width,
are naturally quite small.

\end{document}